# Twin Domains in 111 oriented {CdO/MgO} superlattices: homoepitaxy versus heteroepitaxy

*Ewa Przeździecka, Aleksandra Wierzbicka\*, Abinash Adhikari, Marta A. Chabowska*

Institute of Physics Polish Academy of Sciences, Al. Lotników 32/46, Warsaw, Poland

KEYWORDS: Superlattices, X-ray diffraction, Twin Domains, CdO, MgO, Epitaxy

**Abstract**

The structural properties of (111)-oriented {CdO/MgO} superlattice structures grown on *c*-sapphire and cubic MgO substrates have been studied by high resolution X-ray diffraction. The growth was performed in a plasma-assisted molecular beam epitaxy system. Although both superlattices are (111)-oriented and the {CdO/MgO} structure has 3m symmetry. It was shown that the superlattice on *c*-sapphire consists of misoriented domains, whereas no such domains were observed on (111) MgO. The twin domains are rotated by 180° with respect to each other and by ±30° with respect to the sapphire substrate. We show that the crucial phenomena based on the formation of rotation domains and their number in heteroepitaxy depend fundamentally on the relationship between substrate and epilayer symmetries.

**Introduction**

In epitaxial growth, the substrate and epilayer can differ significantly in crystal structure and lattice parameters (heteroepitaxy) or be the same in both respects (homoepitaxy). Mismatches in translational symmetry lead to specific types of defects, such as dislocations. This well-known phenomenon has been studied extensively, and domain epitaxy provides an explanation for the advanced matching conditions enabled by specific lattice surface rotations. Rotation domains in the epilayer are characterised by identical crystallographic direction along the growth axis, but different in-plane azimuthal orientations. The formation and abundance of these domains in heteroepitaxy depends fundamentally on the symmetry relationship between the substrate and the epilayer, with both rotational and mirror symmetry playing key roles [1]. For example, in an AlN (00.1) epilayer grown on a Si(111) substrate, both the epilayer and the substrate exhibit 3m point symmetry, resulting in the expected single domain growth [2,3]. In a similar case with a GaN(00.1) epilayer on Ge(111), where both materials again have 3m symmetry, the mirror symmetry of the Ge surface allows for equivalent GaN nucleation rotated by ±4°, leading to the formation of two domains [4,5]. Growth conditions and surface treatment can also influence the number of domains observed [6]. For example, in ZnO (00.1) layers grown on (00.1) sapphire, either one or two rotation domains can be observed depending on the conditions [7,8]. Twin domains have also been observed in both CdO and MgO films [9,11]. The presence of twins can significantly affect the physical properties of layers, influencing aspects such as doping [12-15], making the identification of rotation domains essential in semiconductor structures.

CdO thin films have found efficient applications in transparent conducting oxides (TCOs) and solar cells [16]. By combining CdO with MgO, it is possible to produce a ternary alloy where the energy gap can be tuned from yellow (2.1 eV) to deep ultraviolet (7 eV) [17-19]. The creation of {CdO/MgO} superlattices allows precise control of the energy gap of this quasi-ternary material [20]. In this work, we analyse and compare (111)-oriented {MgO/CdO} rocksalt superlattices (SLs) grown on (00.1) *c*-oriented sapphire and (111) MgO substrates. Our aim is to systematically study the planar defects present in oxide superlattices using high resolution X-ray diffraction techniques.

**Methods**

A plasma-assisted molecular beam epitaxy (MBE) system (RIBER Compact 21) was used to grow series of {CdO/MgO} superlattices (SLs) samples on the (00.1) sapphire and (111) MgO substrates. The Cd and Mg fluxes were provided by evaporation of elemental Cd and Mg (6N) from a commercial Knudsen cell. A radio frequency (RF) plasma system was used to generate the active oxygen, with the oxygen gas flow rate controlled by a mass flow controller. Prior to growth, the substrates were chemically etched for 5 minutes in a solution of $H_2SO_4$ : $H_2O_2$ = 1 : 1 to remove surface contaminants. The substrates were then thermally cleaned at approximately 700°C for 5 minutes followed by 30 minutes of O* irradiation to obtain an oxygen terminated surface. O* irradiation was performed at an RF power of approximately 400 W and an oxygen gas flow rate of 3 sccm.

A series of {CdO/MgO} superlattices with CdO sublayer thicknesses ranging from 1 monolayer (ML) to 10 ML, and a constant MgO thickness of 4 ML, were grown at a substrate temperature of 360°C, measured by thermocouple. Growth parameters such as flux, growth temperature, and oxygen flow were consistent for both MgO and $Al_2O_3$ substrates. During the growth, the number of repetitions varied from 21 to 45, depending on the thickness of the sublayers, so that the total thickness of the superlattices was comparable (see Table 1).

| Samples on MgO and c-plane sapphire substrates | | |
|---|---|---|
| **MgO / CdO ML MBE** | **Number of repetitions** | **MgO / CdO thickness XRD (nm)** |
| 4 ML MgO / 1 ML CdO | 45 | 2.25 / 0.5 |
| 4 ML MgO / 2 ML CdO | 40 | 2.25 / 0.8 |
| 4 ML MgO / 4 ML CdO | 32 | 2.25 / 2.2 |
| 4 ML MgO / 6 ML CdO | 27 | 2.25 / 3.17 |
| 4 ML MgO / 10 ML CdO | 21 | 1.73 / 3.6 |

**Table 1.** Number of individual monolayers of {CdO/MgO} superlattices in MBE growth, total number of iterations of the MgO and CdO monolayer process, and interlayer thicknesses obtained from XRD measurements.

The surface morphology was monitored by atomic force microscopy (AFM) and the crystal quality and orientation of the superlattices were characterised by high resolution X-ray diffraction (HRXRD, Panalytical X'Pert Pro MRD). The diffractometer, equipped with a $Cu_{K\alpha 1}$ radiation source ($\lambda$ = 1.54056 Å), had a hybrid 2-bounce Ge (220) monochromator and two detection modes: low resolution (with a Pixcel detector and Soller slits) and high resolution (with a proportional detector and hybrid 3-bounce Ge (220) analyser).

Three types of XRD measurements were carried out using a Multi Research Diffractometer. Firstly, symmetrical (see Figure 1a) and asymmetrical (see Figure 1b) geometries were used to determine accurate values for lattice parameters and lattice distortion of the measured samples. In these geometries the incident and scattered vectors (kin and kout vectors in Figure 1) lie in a plane perpendicular to the sample surface, this plane is called the scattering plane. The main difference between symmetric and asymmetric geometries is the angle formed by the incident ($\omega i$) and diffracted ($\omega e$) rays relative to the sample surface. In symmetrical geometry, the scattering angle and the incident angle are identical ($\omega = \theta$), whereas in asymmetrical geometry they are different:

$$\omega = \theta \pm \alpha \qquad (1)$$

where $\omega$ is the angle between the incident beam and the sample surface, $2\theta$ is the angle between the scattered beam and the incident beam and $\alpha$ is the inclination of the lattice plane relative to the surface (see Figure 1).

The third type of measurement technique used in this work is symmetrical non-coplanar diffraction geometry (Figure 1c) called skew diffraction geometry. In skew geometry, the scattering plane is inclined at an angle $\chi$ relative to the sample surface, rather than being perpendicular to it. In this case, the incident and diffracted Bragg angles of the X-ray beam remain identical ($\omega = \theta$) (see Figure 1c).

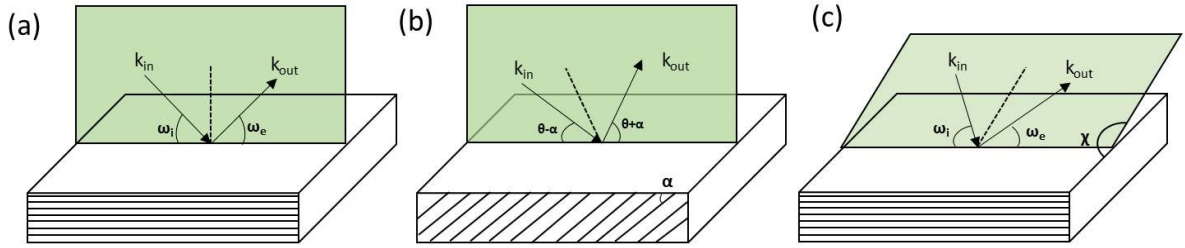

**Figure 1.** Schematic representation of X-ray diffraction geometries. (a) Symmetrical and (b) asymmetrical coplanar X-ray diffraction. The incident and diffracted X-ray beam are in the normal plane perpendicular to the sample surface. (c) Non-coplanar XRD skew geometry, the incident and diffracted X-ray beam is tilted ($\chi$-angle) to the sample surface.

Precise measurements of the absolute position and integrated intensity of different reciprocal lattice points provided insight into the structural properties of the samples. For example, the lattice content, sublayer thickness and strain of the layers could be inferred from the position of a given reciprocal lattice point. By examining the scattered intensity around a reciprocal lattice point, information about strain fields, shape or dislocation density was obtained as it reflects the square Fourier transform of the shape function.

**Results and discussion**

As mentioned above, the growth parameters were chosen so that the total thickness of the superlattice was comparable. This allows direct comparison of the X-ray diffraction results. The low angle resolution $\theta$-$2\theta$ scans shown in Figure 2 confirm the presence of (111) oriented cubic {CdO/MgO} SLs on both *c*-sapphire and (111) MgO substrates. As previously published and confirmed experimentally and theoretically for this type of structure, the lattice parameters of the superlattices change with the change in thickness of the CdO sub-layers at a fixed thickness of the MgO sub-layers [21]. The change in the average lattice constants for SLs is accompanied by a change in the angular positions of the peaks. The signals from (111) structures on both substrates are similar, except that for samples with a smaller thickness of CdO layer, the signal merges with the signal from the MgO substrate (Figure 2a).

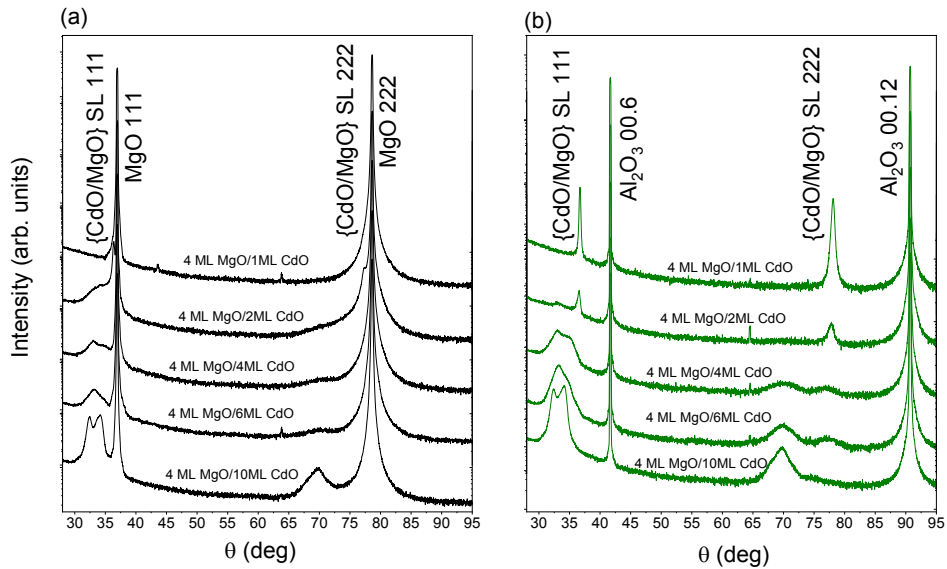

**Figure 2.** θ-2θ scans for a series of (111) oriented {CdO/MgO} superlattices grown on (a) (111) MgO substrates and (b) (00.1) $Al_2O_3$ substrates.

The similar behaviour we observe for {CdO/MgO} SLs grown on $Al_2O_3$ substrates (Figure 2b). The thicker the CdO sublayer in the SL, the more defined the XRD signal from the SL that we observe. In addition, the X-ray beam is more sensitive to differences in the crystallographic lattices of CdO and MgO. In both cases the XRD signal is shifted to smaller Bragg angles, so the $a_\perp$ lattice constants are highest for the thicker CdO sublayer.

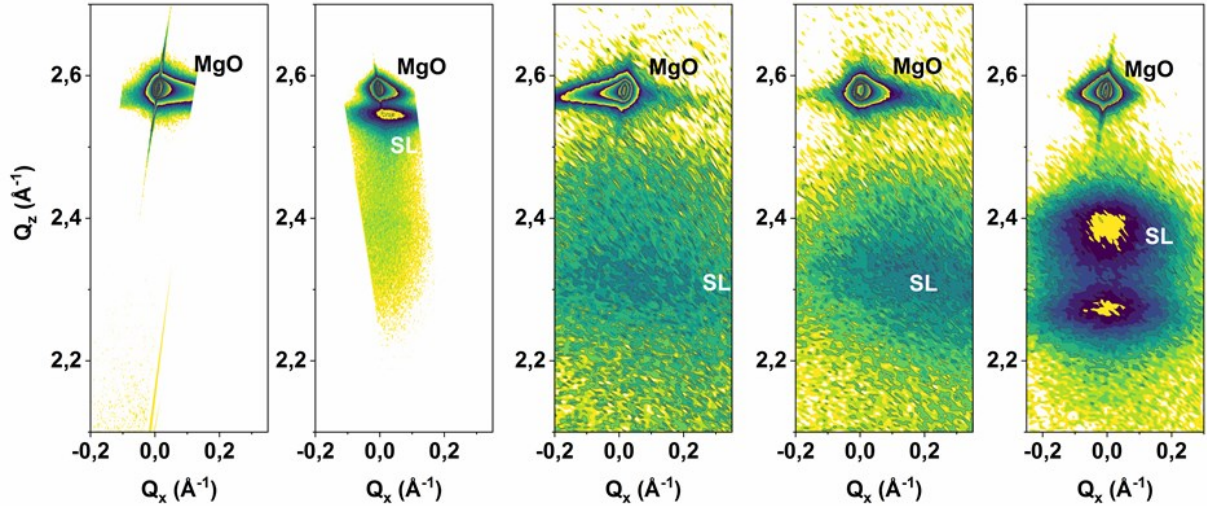

**Figure 3.** High-resolution X-ray diffraction reciprocal space maps of {CdO/MgO} SLs on MgO substrates. This map shows the 111 MgO and the 111 {MgO/CdO} SLs reflections in reciprocal lattice space. In this figure, $Q_x$ and $Q_z$ are lattice vectors in reciprocal lattice space with directions parallel and perpendicular to the substrate surface, respectively.

X-ray diffraction reciprocal space maps (RSM) were measured for further investigation. Figure 3 shows the XRD analysis of the RSM examined in symmetrical geometry measured for a series of samples grown on a MgO substrate. In this case the reciprocal lattice point of the {CdO/MgO} (111) lattice planes is examined. Initially, the strongest signal comes from the MgO substrate. With increasing thickness of the CdO sublayer in SLs we observe the better formation of reciprocal lattice points of {CdO/MgO} SL. Increasing the thickness of the CdO layer causes a shift of the diffracted signal from the (111) MgO lattice point to smaller values of the $Q_z$ vector. RSM analysis allowed us to determine the lattice parameters of the SL and the substrate.

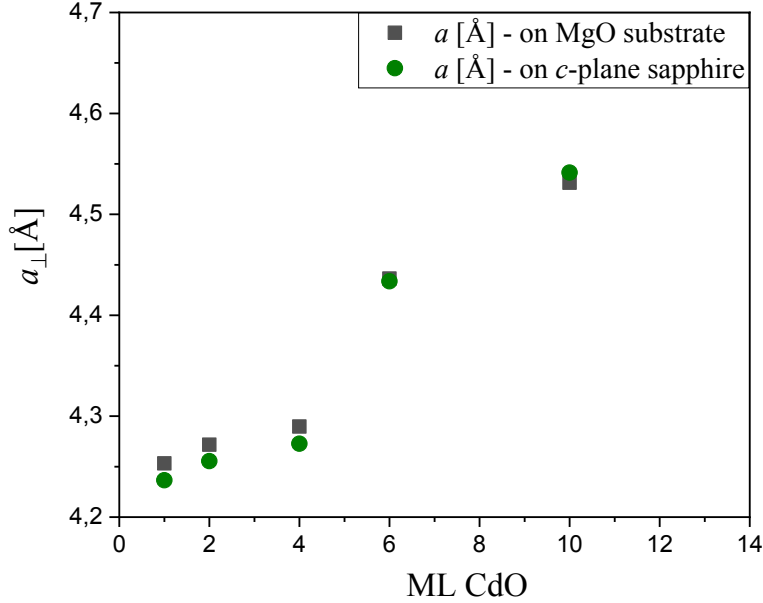

**Figure 4.** Dependence of the lattice parameters of {CdO/MgO} SLs on the number of CdO monolayers in the superlattice calculated from XRD reciprocal space maps for structures on MgO and $Al_2O_3$ substrates.

The lattice parameter $a_\perp$ of the MgO substrate is the same for all samples ($a_{MgO}$ = 4.2046 Å). For the superlattice, we observe an increase in the lattice parameter $a_\perp$ with increasing thickness of the CdO layer (see Figure 4). The lattice parameters for series of samples grown on $Al_2O_3$ substrates were also calculated from RSMs. All the results obtained for 2 series of samples are shown in Figure 4. The green circles are the lattice parameters of SLs grown on *c*-plane sapphire and the black squares are the lattice parameters of SLs grown on MgO substrates. These values are compared with the theoretical values calculated in our previous work (Figure 9 in reference [21]). The experimental data are comparable regardless of the type of substrate used.

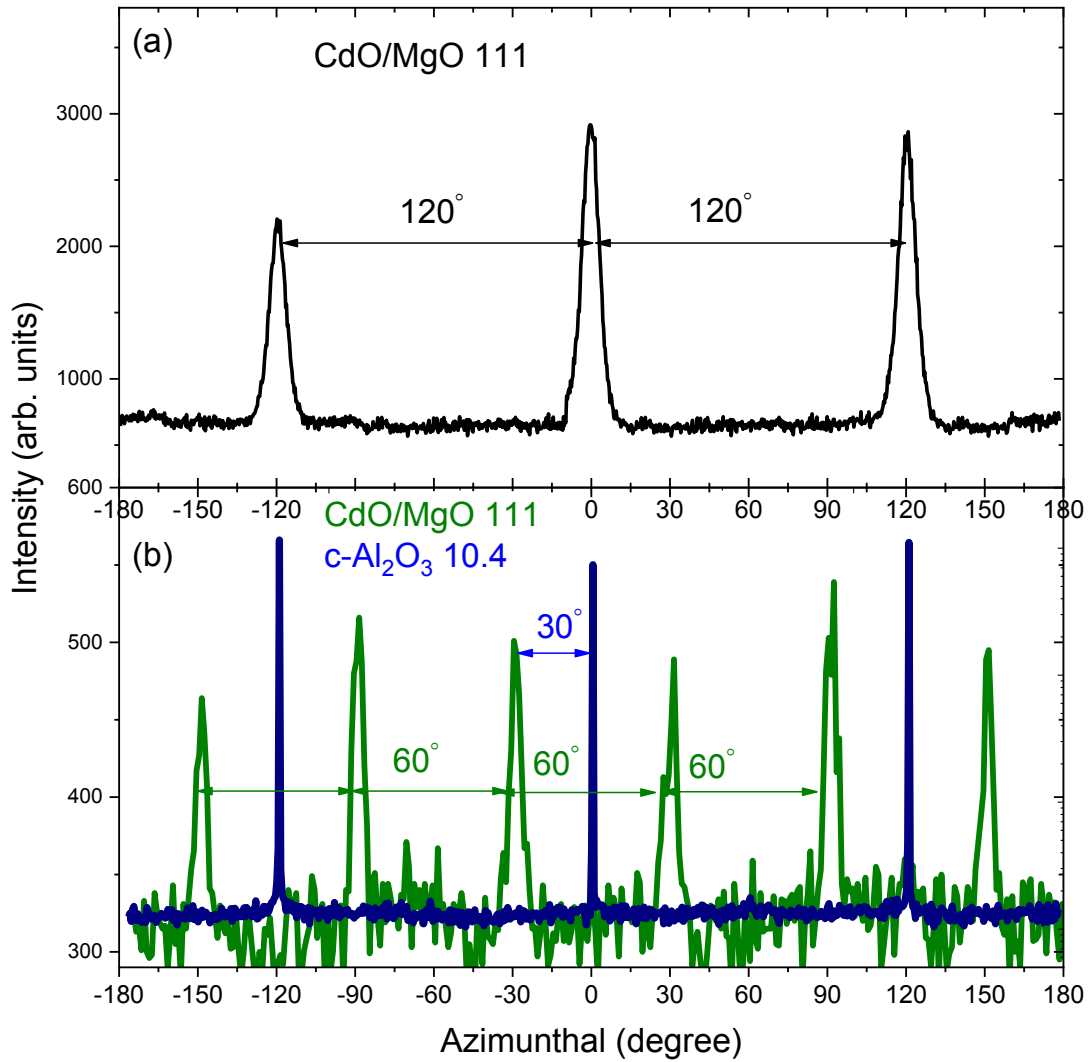

**Figure 5.** X-ray diffraction $\varphi$-scans measured in skew geometry ($\chi \sim 54°$) for (a) 111 {CdO/MgO} SL reflection grown on MgO substrate and (b) for the 10.4 $Al_2O_3$ reflection substrate and 111 {CdO/MgO} SL reflection grown on $Al_2O_3$ substrate. Data are shown for samples with 10 ML of CdO.

In the next step, the third geometry of XRD measurements is presented. We measure symmetrical scans for lattice planes rotated by $\chi$ angle for the surface sample (compare with Figure 1 c). First, the symmetrical 111 {CdO/MgO} peak was studied for samples grown on MgO substrates and then on c-sapphire substrates. The diffracted signal from $Al_2O_3$ was also recorded. Since the planes of the substrate and the growing structures typically have different periodicities and thus different Bragg angles, they can be recorded independently. From the φ - scan, it is possible to record the variation of the azimuthal orientation around the substrate normal and reveal the in-plane rotation of the domains. The in-plane orientation relationship between {CdO/MgO}, MgO and the *c*-sapphire substrates was determined using the φ-scan method. The results are shown in Figure 5. It can be seen that the φ scan for {CdO/MgO} on MgO substrates gives three peaks, confirming the epitaxial growth. The presence of these reflections is attributed to the threefold symmetry of the {CdO/MgO} film growth. Furthermore, only three signals originating from SLs are visible on the the MgO substrate, whereas six diffracted signals are observed on *c*-sapphire (Figure 3b). These six peaks are associated with two families of ±180° twisted triangular domains.

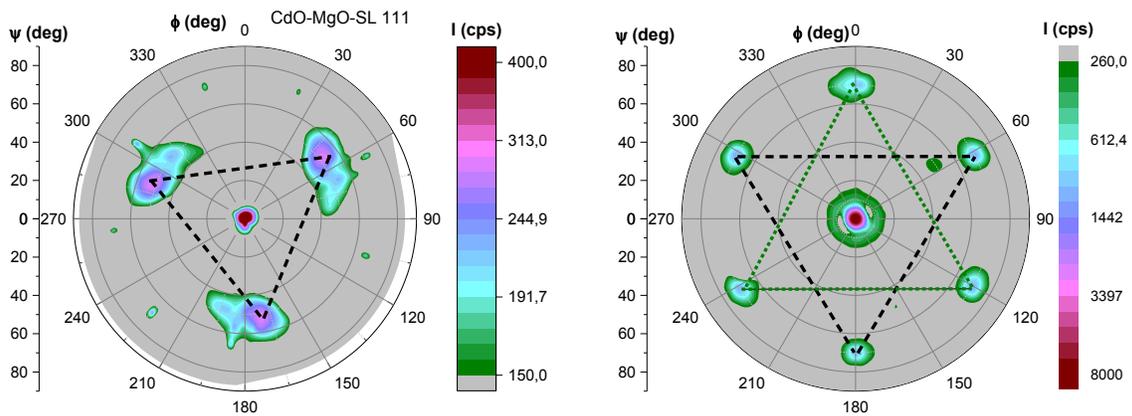

**Figure 6.** The XRD pole figures of 111 {CdO/MgO} SLs reflections on (a) MgO substrate (b) on *c*-$Al_2O_3$ substrate.

The presence of these two families is also confirmed in XRD pole figures (Figure 6b), in contrast to the single domain visible in Figure 6a on MgO substrates. The signals from the SLs are not sharp because the signals from the satellite peaks are also visible (fringes coming from SL). In the case of growth on sapphire, the following epitaxial relationships are expected: $[1100]Al_2O_3 \parallel [110)]\{MgO/CdO\}$, $[1120]Al_2O_3 \parallel [112]\{MgO/CdO\}$, and $[0001]Al_2O_3 \parallel [111]\{MgO/CdO\}$ [10]. As can be seen in Figure 5b in our case the {CdO/MgO} signal is twisted by 30° to the expected in plain relation. This symmetry breaking allows twins to form, although they should not occur for the same 3m symmetry of the substrate used and the grown layers. According to previous reports for MgO layers grown at relatively high temperatures (700°C), the existence of the spinel $MgOAl_2O_3$ existence should be considered [19]. The presence of additional, very thin spinel layers can influence on the formation of twin domains. However, in our case a relatively low growth temperature (360°C) was used, so it seems that $MgOAl_2O_3$ should not be formed – which is also confirmed by the previously published Transmission Electron Microscopy (TEM) scans [20]. Therefore, we believe that the breaking of the point symmetry by the 30° rotation in the growth plane is responsible for the formation of twins in these SL structures (Figure 7).

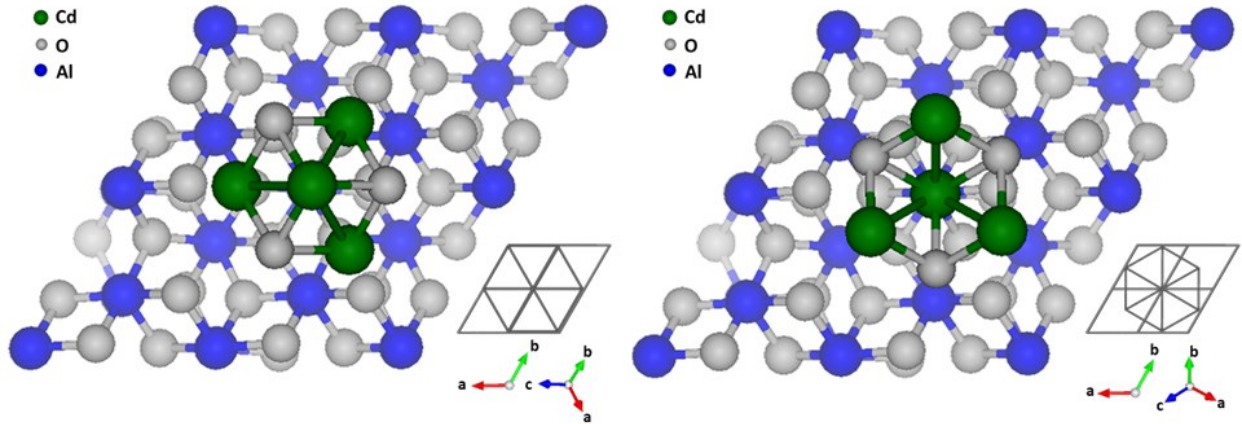

**Figure 7.** Atomic model of the (111)CdO/(00.1)Al$_2$O$_3$ structure: in-plane projection (a) without rotation (b) with 30° rotation. The blue and gray spheres represent Al and O atoms, respectively, and the green spheres represent Cd atoms. The structures were generated using ViewerLite 5.0 software.

**Conclusions**

Reducing defects and understanding their origin in semiconductor crystals and structures plays an important role in improving the performance of future optoelectronic devices. In this paper, an advanced XRD technique has been used to demonstrate the presence of domain twins in {CdO/MgO} superlattices. For this purpose, we compared two series of SLs structures grown in the (111) direction, which at first sight appear identical. By using combinations of typical symmetrical and asymmetrical XRD geometries with the symmetrical XRD skew geometry, we were able to show the presence of 180 degree rotated twins on the *c*-sapphire substrate. The reason for the formation of these domains is the fact that the *c*-Al$_2$O$_3$ plane is twisted by 30 degrees relative to the (111) CdO or MgO plane.

Corresponding Author

*Aleksandra Wierzbicka, wierzbicka@ifpan.edu.pl

**Author Contributions**

Ewa Przeździecka: Writing – original draft, Investigation, Data curation, Supervision, Funding acquisition, Project administration, Conceptualization. Aleksandra Wierzbicka: Writing – original draft (part about XRD investigatons) review & editing, Methodology, Investigation, Resources, Conceptualization. Abinash Adhikari: Investigation, Methodology, Formal analysis, Visualization. Marta A. Chabowska: Formal analysis, Visualization, original draft review & editing.

**Notes**

The authors declare no competing financial interest.

**Acknowledgments**

This work was supported in part by the Polish National Science Center, Grant No. 2021/41/B/ST5/00216.